\newif \iftwocol
\twocoltrue
\iftwocol
\documentclass[10pt,conference]{IEEEtran}
\else
\documentclass[11pt,onecolumn]{IEEEtran}
\fi

\usepackage{psfrag}

\usepackage{amsmath}
\usepackage{amsthm}
\usepackage{amssymb}
\usepackage{latexsym}
\usepackage{cite}
\usepackage{graphicx}
\usepackage{tabularx}
\usepackage{url}
\usepackage{booktabs}
\usepackage{subfigure}
\usepackage{color}

\newcommand{\BIT}{\begin{itemize}}
\newcommand{\EIT}{\end{itemize}}
\newcommand{\BEN}{\begin{enumerate}}
\newcommand{\EEN}{\end{enumerate}}

\newcommand{\Pb}{\mathbf{P}}

\newcommand{\C}{\mathcal{C}}
\newcommand{\Gb}{\mathbf{G}}

\newcommand{\U}{\mathcal{U}}

\newcommand{\Vb}{\mathbf{V}}

\newcommand{\Hb}{\mathbf{H}}
\newcommand{\Yb}{\mathbf{Y}}

\newcommand{\xb}{\mathbf{x}}

\newcommand{\Xb}{\mathbf{X}}

\newcommand{\Zb}{\mathbf{Z}}

\newcommand{\Tb}{\mathbf{T}}

\newcommand{\wb}{\mathbf{w}}

\newcommand{\Xm}{\mathbf{X}}
\newcommand{\Id}{\mathbf{I}}
\newcommand{\Sm}{\mathbf{S}}
\newcommand{\Ym}{\mathbf{Y}}

\newcommand{\Gm}{\mathbf{G}}
\newcommand{\Pm}{\mathbf{P}}
\newcommand{\xv}{\mathbf{x}}
\newcommand{\sv}{\mathbf{s}}
\newcommand{\uv}{\mathbf{u}}
\newcommand{\Cc}{\mathcal{C}}

\newcommand{\NoN}{\nonumber}

\newtheorem{remark}{Remark}

\newcommand{\executeiffilenewer}[3]{
\ifnum \pdfstrcmp{\pdffilemoddate{#1}}{\pdffilemoddate{#2}}>0
{\immediate\write18{#3}}
\fi
}

\newcommand{\Ic}{\mathcal{I}}
\newcommand{\Sc}{\mathcal{S}}
\newcommand{\Pc}{\mathcal{P}}
\newcommand{\zerov}{\mathbf{0}}

\newfont{\bb}{msbm10 scaled 1100}
\newcommand{\CC}{\mbox{\bb C}}
\newcommand{\PP}{\mbox{\bb P}}

\newcommand{\ZZ}{\mbox{\bb Z}}

\begin{document}
\title{A Nonlinear Approach to Interference Alignment}
\author{\IEEEauthorblockN{Peyman Razaghi and Giuseppe Caire}
\IEEEauthorblockA{Department of Electrical and Computer Engineering\\
University of Southern California\\
Los Angeles, California, USA\\
emails: {\tt \{razaghi,caire\}@usc.edu}}
}


\maketitle

\begin{abstract}
Cadambe and Jafar (CJ) alignment strategy for the $K$-user scalar frequency-selective fading Gaussian channel, with encoding over blocks of $2n+1$ random channel coefficients (subcarriers) is considered. The linear zero-forcing (LZF) strategy is compared with a novel approach based on lattice alignment and lattice decoding (LD). Despite both LZF and LD achieve the same degrees of freedom, it is shown that LD can achieve very significant improvements in terms of error rates at practical SNRs with respect to the conventional LZF proposed in the literature. We also show that these gains are realized provided that channel gains are controlled to be near constant, for example, by means of power control and opportunistic carrier and user selection strategies. In presence of relatively-small variations in the normalized channel coefficient amplitudes, CJ alignment strategy yields very disappointing results at finite SNRs, and the gain of LD over ZLF significantly reduces. In light of these results, the practical applicability of CJ alignment scheme remains questionable, in particular for Rayleigh fading channels, where channel inversion power control yields to unbounded average transmit power.
\end{abstract} 

\section{Introduction}
``Everyone gets half the cake'' is the surprising promise of interference alignment, as introduced by Cadambe and Jafar \cite{cadambe_jafar} for a $K$-user Gaussian interference channel with random coefficients. Interference alignment is a linear precoding strategy that forces the interfering signals at each receiver $k$ to span a
subspace $\Ic_k$ of the receiver signal space,  such that the desired signal can be transmitted in a subspace $\Sc_k$ with $\Ic_k \cup \Sc_k = \{\zerov\}$. Therefore, each receiver $k$ can remove interference completely by a linear zero-forcing projection on the orthogonal complement $\Ic_k^\perp$. With enough richness in the channel coefficients, in the form of jointly distributed random variables drawn from a continuous distribution (e.g., as arising from time and/or frequency selective fading channels) \cite{cadambe_jafar} shows that by encoding over a large block of $N$ channel uses, the limit of $K/2$ degrees of freedom is achievable, i.e., $\lim_{N \rightarrow \infty} {\rm dim}(\Sc_k)/N = 1/2$ for all $k$. With such a strategy in place, the entire group of interfering transmitters  at each destination appears collectively as a single source of interference, consuming only half the total degrees of freedom (dimensions) and leaving another half available at each user.

In this paper, we put this idea to test by investigating practical decoding strategies for interference alignment in practical signal-to-noise ratios (SNR). We focus on the parallel channel single-antenna scenario introduced in \cite{cadambe_jafar} for i.i.d. random subchannel coefficients, e.g., OFDM with frequency-selective fading. We find that despite the promise of the degrees-of-freedom analysis,  the quality of the effective channel with interference aiglnment and linear zero-forcing (LZF) interference removal is generally not quite acceptable at practical finite SNRs. This is especially true for channel coefficients with non-constant amplitudes (as in a frequency-selective channel).

The major limiting factor in finite-SNR performance of Cadambe and Jafar (CJ) interference alignment pertains to channel inversion operations at the transmit side precoding, and zero-forcing interference cancelation at the receive side. In order to improve upon the CJ strategy, in this work we consider {\em discrete} alignment of signal sets in addition to the CJ alignment of signal subspaces.
The core idea is to obtain an equivalent MIMO channel suitable to  more efficient non-linear decoding strategies. In order to do so, it is essential that the superposition of interference signals not only span a low-dimensional subspace, but also appear as points in a discrete lattice, closed with respect to addition. 
The advantage of discrete lattice alignment is that it allows to recast the channel observed by each receiver as a MIMO channel where the desired signals occupy approximately half of the signal space dimensions, and the {\em sum} of interfering signals spans the other half.  This allows to apply well-understood Lattice Decoding (LD) strategies such as sphere decoding and variations thereof (e.g., \cite{agrell_eriksson,damen_gamal_caire1,viterbo_boutros}), in order to decode the intended signal along with the superposition of interfering signals. 

We show that by exploiting the discrete nature of the transmitted signals, this nonlinear decoding approach yields a surprising improvement compared to linear alignment with zero forcing, provided that the dynamic variations of channel coefficient amplitudes are controlled. Channel dynamics have degrading effects on the performance largely due to the transmit-side inversive precoding. However, suitable carrier pairing in time and frequency dimensions, and opportunistic user selection strategies like those in \cite{viswanath_tse_laroia} could be used as powerful means to achieve near constant channel amplitudes.


The rest of the paper is organized as follows. Section~\ref{sec:classic} gives a summary of the CJ interference alignment scheme for single-antenna, parallel fading channels. The idea of discrete alignment and the MIMO interpretation of interference channel with alignment is introduced Section~\ref{sec:nonlinear}, and Section~\ref{sec:sim} presents performance results and simulations.
Section \ref{sec:con} concludes the paper with a few final remarks.  

\section{Linear Subspace Alignment\label{sec:classic}}

We focus on the $K = 3$ user case, with precoding block length $N = 2n + 1$ for some integer $n \geq 1$,
as in \cite{cadambe_jafar}.
A channel use of this channel is given by
\begin{align}
Y_1^t&=H_{11}^tX_1^t+H_{12}^tX_2^t+H_{13}^tX_3^t+Z_1^t\NoN\\
Y_2^t&=H_{21}^tX_1^t+H_{22}^tX_2^t+H_{23}^tX_3^t+Z_2^t\NoN\\
Y_3^t&=H_{31}^tX_1^t+H_{32}^tX_2^t+H_{33}^tX_3^t+Z_3^t,\NoN
\end{align}
where $\{H_{ij}^t\}$ denote the channel coefficient, and $\{Y_i^t,X_i^t,Z_i^t\}$ represent the received symbols, the transmitted symbols,
and the noise samples at channel use $t$, respectively, for $i,j=1,2,3$.
The CJ strategy is a precoding scheme across a block of dimensions (e.g., subcarriers in an OFDM channel).
In a block of length $2n+1$ channel uses,
user 1 encodes a vector $\Xb_1$ of length $n+1$ symbols using a precoding matrix
 $\Vb_1 \in \CC^{(2n+1)\times (n+1)}$, while users 2 and 3 encode $\Xb_2$ and $\Xb_3$ of length $n$ using precoding
 matrices $\Vb_2, \Vb_3 \in \CC^{(2n+1)\times n}$, respectively.
 Notice that the encoding scheme is not completely symmetric. However, symmetry can be achieved on average,
 by a rotating scheduling among users over successive blocks.

The equivalent block channel can be described as
\begin{subequations}
\begin{align}
\Yb_1&=\Hb_{11}\Vb_1\Xb_1+\Hb_{12}\Vb_2\Xb_2+\Hb_{13}\Vb_3\Xb_3+\Zb_1\\
\Yb_2&=\Hb_{21}\Vb_1\Xb_1+\Hb_{22}\Vb_2\Xb_2+\Hb_{23}\Vb_3\Xb_3+\Zb_2\\
\Yb_3&=\Hb_{31}\Vb_1\Xb_1+\Hb_{32}\Vb_2\Xb_2+\Hb_{33}\Vb_3\Xb_3+\Zb_3,
\end{align}
\end{subequations}
where $\Yb_i,\Zb_i$ represent the received and noise vectors of length $2n+1$, $\Xb_i$ represent the users' data vectors,
and $\Hb_{ij}$ denote the diagonal channel matrix of size $(2n+1)\times(2n+1)$, for $i,j=1,2,3$.

The precoding  matrices $\Vb_1,\Vb_2,\Vb_3$  are designed such that the interfering signals occupy a common subspace.
This {\em alignment} is achieved in \cite{cadambe_jafar} through the following design:
\begin{align}
\begin{array}{rcl}
\Vb_1&=&\bigl[\begin{array}{cccc}\wb & \Tb\wb & \ldots & \Tb^n\wb\end{array}\bigr]\\
\Vb_2&=&\Hb_{32}^{-1}\Hb_{31}\Vb_1\Pb_2\\
\Vb_3&=&\Hb_{23}^{-1}\Hb_{21}\Vb_1\Pb_3
\end{array},
&
\wb&=\left[\begin{array}{c}
1\\
1\\
\vdots\\
1
\end{array}\right]^{(2n+1)\times 1}\label{eq:pcoder1}
\end{align}
and,
\iftwocol
\begin{align}
\Tb&=\bigl(\Hb_{31}^{-1}\Hb_{32}\bigr)\bigl(\Hb_{12}^{-1}\Hb_{13}\bigr)\bigl(\Hb_{23}^{-1}\Hb_{21}\bigr),\label{eq:pcoder4}\\
\Pb_2^{(n+1)\times n}&=\left[\begin{array}{c}\mathbf{0}\\\mathbf{I}_n\end{array}\right], \NoN \Pb_3^{(n+1)\times n}=\left[\begin{array}{c}\mathbf{I}_n\\\mathbf{0}\end{array}\right]. \NoN
\end{align}
\else
\begin{align}
\Tb&=\bigl(\Hb_{31}^{-1}\Hb_{32}\bigr)\bigl(\Hb_{12}^{-1}\Hb_{13}\bigr)\bigl(\Hb_{23}^{-1}\Hb_{21}\bigr), & \Pb_2^{(n+1)\times n}&=\left[\begin{array}{c}\mathbf{0}\\\mathbf{I}_n\end{array}\right], & \Pb_3^{(n+1)\times n}&=\left[\begin{array}{c}\mathbf{I}_n\\\mathbf{0}\end{array}\right], & \wb&=\left[\begin{array}{c}
1\\
1\\
\vdots\\
1
\end{array}\right]^{(2n+1)\times 1}\label{eq:pcoder4}.
\end{align}
\fi
It is fairly straightforward to check that with the above precoding design, we have
\begin{align}\label{eq:aligned}
\Hb_{12}\Vb_2&\overset{(a)}=\Hb_{13}\Vb_3,\NoN\\
\Hb_{23}\Vb_3&=\Hb_{21}\Vb_1\Pb_3,\NoN\\
\Hb_{32}\Vb_2&=\Hb_{31}\Vb_1\Pb_2,
\end{align}
where (a) follows since by the structure of $\Tb$ and the shift property of $\Pb_2$ and $\Pb_3$, we have $\Vb_1\Pb_2=\Tb\Vb_1\Pb_3$. As a consequence, the two interfering signals received at user $i$ are confined to a common subspace spanned by the columns of
$\Hb_{12}\Vb_2$, $\Hb_{21}\Vb_1$, and $\Hb_{31}\Vb_1$ for $i=1,2,3$, respectively. This is readily seen if we rewrite \eqref{eq:blkch} as
\begin{subequations}\label{eq:alignedch}
\begin{align}
\Yb_1&=\Hb_{11}\Vb_1\Xb_1+\Hb_{12}\Vb_2(\Xb_2+\Xb_3)+\Zb_1\\
\Yb_2&=\Hb_{22}\Vb_2\Xb_2+\Hb_{21}\Vb_1(\Xb_1+\Pb_3\Xb_3)+\Zb_2\\
\Yb_3&=\Hb_{33}\Vb_3\Xb_3+\Hb_{31}\Vb_1(\Xb_1+\Pb_2\Xb_2)+\Zb_3.
\end{align}
\end{subequations}

The aligned structure of the interference vectors allows the decoders to remove
interference by LZF, i.e., by  projecting the received vector onto  the orthogonal complement of the interference subspace.

\begin{remark}\label{rem:1}
For successful zero forcing, we also need the interference subspace to be linearly independent of the
signal subspace, i.e., $[\begin{array}{cc}\Hb_{11}\Vb_1 & \Hb_{12}\Vb_2\end{array}]$, $[\begin{array}{cc}\Hb_{22}\Vb_2 & \Hb_{21}\Vb_1\end{array}]$,  and $[\begin{array}{cc}\Hb_{33}\Vb_3 & \Hb_{31}\Vb_1\end{array}]$ in \eqref{eq:alignedch} are full rank.  In \cite{cadambe_jafar}, it is shown that this rank constraint is almost surely satisfied for channel coefficients  drawn from a non-degenerate
continuous distribution (i.e., a distribution for which no coefficient is a deterministic function of other coefficients).
Channel coefficients are further constrained in \cite{cadambe_jafar} to satisfy $a \leq |H_{ij}^t| \leq b$, for some $0 < a \leq b < \infty$.
(see \cite[Section~II]{cadambe_jafar}.)
This magnitude constraint, in particular, rules out direct application of this alignment strategy in a Rayleigh fading environment, for which channel inversion yields unbounded average transmit power.
However, this strictly-bounded constraint can be enforced, in practice, by power control and opportunistic user selection. For example, we may assume that each subcarrier
$t$ in an OFDM system is pre-multiplied by a power control function $\sqrt{\Pc_i^t}$ such that for all $t$ and all users $i$, $|H_{ji}^t\sqrt{\Pc_i^t}|$ is bounded, for all $j$. Notice also that simple channel inversion does not accomplishes constant channel amplitudes, since a single transmitter
must equalize the power of all its $K$ outgoing links simultaneously. However, we may imagine some form of opportunistic subcarrier pairing across multiple users \cite{viswanath_tse_laroia}, such that
groups of $2n+1$ subcarriers are chosen in order to have roughly the same amplitudes. By doing so,
we have that $\lvert H_{i1}^t\rvert \approx \lvert H_{i2}^t\rvert  \approx \lvert H_{i3}^t\rvert$ for all $i = 1,2,3$, and the power control command $\Pc_i^t$
is used to equalize amplitudes across the three transmitters.
\end{remark}
    %
    %
    %
    %
    %
    %

\section{Discrete Alignment with Nonlinear Decoding\label{sec:nonlinear}}
Qualitatively, the key element in interference alignment is constricting the ``expansion'' resulting from linear superposition
by confining the interfering signals to a common subspace.
For discrete signals, however, we can prevent this ``expansion'' if the codewords form a discrete additive group.
For example, consider a lattice $\Lambda=\{\Gb\xb\vert\xb\in\mathbb{Z}^n\}$, where $\Gb$ is a full rank generator matrix.
Summation of two codewords $\xb_1,\xb_2\in\Lambda$ gives another codeword $\xb_1+\xb_2\in\Lambda$.

%

Consider for example the following interference channel
\begin{align}
Y_1&=g_1 X_1+(X_2+X_3)+Z_1\NoN\\
Y_2&=g_2 X_2+(X_1+X_3)+Z_2\NoN\\
Y_3&=g_3 X_3+(X_1+X_2)+Z_3,\label{eq:ex1}
\end{align}
where $X_1,X_2,X_3$ are chosen from a rectangular QAM constellation $\C$, given as a subset of the complex integer lattice
$\ZZ[j]$.   
Receiver $i$ attempts to recover the desired symbol $X_i$ by treating $X_j + X_k$ (for $j,k \neq i$) as points in an expanded
constellation, obtained as the sum set $\C' = \{x \in \CC : x = y+z, (y,z) \in \C^2\}$. 
In general, if symbols are uniform over $\C$, the resulting extended constellation $\C'$ is used with non-uniform probability; however, for the time being, we neglect this fact and consider maximum-likelihood (ML) decoding (assuming uniform prior probability for $X_i \in \C$ and $X_j + X_k \in \C'$).
A more general one-dimensional strategy is proposed in \cite{motahari_gharan} and is shown to achieve the maximum
asymptotic degrees of freedom for a scalar interference channel with deterministic (fixed) coefficients.
See also \cite{jafarian_jose_vishwanath,sridharan_jafarian_vishwanath} where  alignment strategies using lattice codes
are presented for an interference channel with (up-to-a-scaling-factor) rational channel coefficients.

\subsection{Lattice Decoding}
We can combine the CJ ``linear space'' alignment strategy with the above (algebraic) lattice alignment idea.
Let the data symbol vectors $\Xb_1,\Xb_2,\Xb_3$ take on values in Cartesian product subsets of
the cubic lattice with uniform probability, i.e.,
we let $\Xb_1 \in \C^{n+1}$, $\Xb_2 \in \C^n$ and $\Xb_3 \in \C^n$, with $\C \subset \ZZ[j]$.
Therefore, the interference terms $\Xb_2+\Xb_3$, $\Xb_1+\Pb_3\Xb_3$ and $\Xb_1+\Pb_2\Xb_2$
in \eqref{eq:alignedch} are vectors drawn from the complex cubic lattices $\ZZ^{n}[j]$ and $\ZZ^{n+1}[j]$, respectively.

Rewriting \eqref{eq:alignedch} as
\begin{subequations}\label{eq:mimo}
\begin{align}
\Yb_1&=\bigl[\begin{array}{cc}\Hb_{11}\Vb_1 & \Hb_{12}\Vb_2\end{array}\bigr] \left[\begin{array}{c}\Xb_1 \\ \Xb_2+\Xb_3 \end{array}\right]+\Zb_1\\
&:=\Gb_1\tilde{\Xb}_1+\Zb_1\NoN\\
\Yb_2&=\bigl[\begin{array}{cc}\Hb_{22}\Vb_2 &\Hb_{21}\Vb_1\end{array}\bigr]\left[\begin{array}{c}\Xb_2\\\Xb_1+\Pb_3\Xb_3\end{array}\right]
+\Zb_2\NoN\\
&:=\Gb_2\tilde{\Xb}_2+\Zb_2\NoN\\
\Yb_3&=\bigl[\begin{array}{cc}\Hb_{33}\Vb_3& \Hb_{31}\Vb_1\end{array}\bigr]\left[\begin{array}{c}\Xb_3\\\Xb_1+\Pb_2\Xb_2\end{array}\right]+\Zb_3\NoN\\
&:=\Gb_3\tilde{\Xb}_3+\Zb_3,\NoN
\end{align}
\end{subequations}
user 1 can decode $\Xb_1$ along with {\em interference sum} $\Xb_2+\Xb_3$.  Similarly, user 2 and user 3 decode their intended messages $\Xb_2$ and $\Xb_3$ along with a shifted sum of interference $\Xb_1+\Pb_3\Xb_3$ and $\Xb_1+\Pb_2\Xb_2$,
respectively.

For Gaussian noise, ML decoding of $\tilde{\Xb}_1,\tilde{\Xb}_2,\tilde{\Xb}_3$ amounts to minimum square Euclidean distance decoding. For example, for user 1 this becomes
\[ \widehat{\xb}_1 = \arg\min_{\xb \in \C^{n+1} \times (\C')^n} \;\; \| \Yb_1 - \Gb_1 \xb \|^2. \]
Then, the decoder retrieve the first $n+1$ components of $\widehat{\xb}_1$ as the decision on the desired symbols $\widehat{\Xb}_1$.
This minimum distance decoding is identical to the MIMO decoding in a $(2n+1)\times(2n+1)$  AWGN MIMO channel,
for which a vast amount of research exists.  In particular, since $\Gb_1,\Gb_2,\Gb_3$ are full rank (at least with random channel coefficients), an efficient strategy for minimum distance search consists of ignoring the finite constellation constraints and search over the entire cubic lattice
$\xb \in \ZZ^{2n+1}[j]$, using well-known closest lattice point search algorithms, commonly known as
``sphere decoding'' (see  \cite{damen_gamal_caire1,agrell_eriksson,viterbo_boutros} and references therein).

\subsection{Linear vs Nonlinear Interference Cancellation}
Consider user 1:
\begin{eqnarray}
\Yb_1 &=& \Hb_{11} \Vb_1 \Xb_1 + \Hb_{12} \Vb_2 (\Xb_2 + \Xb_3) + \Zb_1 \nonumber \\
& = & \Gb_{11} \Xb_1 + \Gb_{12} (\Xb_2 + \Xb_3) + \Zb_1.
\end{eqnarray}
Consider ML decoding by treating the interference $\Xb_2 + \Xb_3$ as arbitrary unknown vectors.
This is equivalent to assume $\Gb_{12} (\Xb_2 + \Xb_3) = \Sm$, where $\Sm$ is an arbitrary unknown vector with the constraint
$\Sm \in {\rm Span}(\Gm_{12})$, i.e., $\Sm$ is any vector in the linear space spanned by the columns of $\Gm_{12}$.

Under such assumption, the standard approach consists of the so-called Generalized Likelihood Ratio Test (GLRT):
\begin{equation}
\widehat{\xv}_1 = \arg \max_{\xv_1 \in \Cc^{n+1}} \;\; \max_{\sv \in \mbox{Span}(\Gm_{12})} \; p_{Y_1|X_1,S}(\Ym_1|\xv_1, \sv),
\end{equation}
which is equivalent to:
\begin{equation}
\widehat{\xv}_1 = \arg \min_{\xv_1 \in \Cc^{n+1}} \;\; \min_{\sv \in \mbox{Span}(\Gm_{12})} \; \left \| \Ym_1 - \Gm_{11} \xv_1 - \sv \right \|^2.
\end{equation}
Writing $\sv = \Gm_{12} \uv$, with $\uv \in \CC^n$, the inner minimization is a standard Least-Squares problem, whose solution is given by
\begin{equation}
\widehat{\uv} = \left ( \Gm_{12}^\dagger \Gm_{12} \right )^{-1} \Gm_{12}^\dagger \left ( \Ym_1 - \Gm_{11} \xv_1 \right ).
\end{equation}
Replacing this into the objective function of the outer minimization, we obtain
\begin{eqnarray}
\widehat{\xv}_1
& = & \arg \min_{\xv_1 \in \Cc^{n+1}} \;\; \left \| \Pm^\perp_{12} \left ( \Ym_1 - \Gm_{11} \xv_1 \right )\right \|^2,
\end{eqnarray}
where
\[ \Pm^\perp_{12} = \left ( \Id - \Gm_{12} \left ( \Gm_{12}^\dagger \Gm_{12} \right )^{-1} \Gm_{12}^\dagger\right ) \]
is the orthogonal projector onto the orthogonal complement of the interference space Span$(\Gm_{12})$.
This projection corresponds to the familiar linear ZF receiver assumed in Cadambe-Jafar.
Therefore, not surprisingly, the linear ZF receiver is the GLRT receiver assuming an arbitrary interference vector
whose only constraint is to lie in a certain subspace.

Now, we assume that the symbol vectors $\Xm_1, \Xm_2, \Xm_3$ take on values in some discrete sets
with uniform probability.  The MAP decoder for $\Xm_1$ takes on the form
\begin{align}
\widehat{\xv}_1
& = \arg \max_{\xv_1 \in \Cc^{n+1}} \; \PP(\Xm_1 = \xv_1 | \Ym_1) \nonumber \\
& = \arg \max_{\xv_1 \in \Cc^{n+1}} \; \frac{p_{Y_1|X_1}(\Ym_1|\xv_1)}{ \sum_{\xv'_1 \in \Cc^{n+1}} p_{Y_1|X_1}(\Ym_1|\xv'_1)}  \nonumber \\
& = \arg \max_{\xv_1 \in \Cc^{n+1}} \NoN\\&\quad \frac{\sum_{\xv_2,\xv_3} \exp \left ( - \left \| \Ym_1 - \Gm_{11} \xv_1 - \Gm_{12} (\xv_2 + \xv_3) \right \|^2 \right )}
{\sum_{\xv'_1 \in \Cc^{n+1}} p_{Y_1|X_1}(\Ym_1|\xv'_1)}.\NoN
\end{align}
Taking the log and neglecting irrelevant terms, we have
\begin{align}
\widehat{\xv}_1
&=  \arg \max_{\xv_1 \in \Cc^{n+1}}\NoN\\
 &\quad \; \log \left ( \sum_{\xv_2,\xv_3} \exp \left ( - \left \| \Ym_1 - \Gm_{11} \xv_1 - \Gm_{12} (\xv_2 + \xv_3) \right \|^2 \right ) \right ).\NoN
\end{align}
At this point, we notice that for sufficiently large SNR the log-sum of exponential terms is dominated by the largest term, which corresponds to the smallest distance.
Therefore, we arrive at the approximated MAP decoder
\begin{equation}
\widehat{\xv}_1
=  \arg \min_{\xv_1 \in \Cc^{n+1}} \; \min_{\xv_2,\xv_3} \; \left \| \Ym_1 - \Gm_{11} \xv_1 - \Gm_{12} (\xv_2 + \xv_3) \right \|^2,\NoN
\end{equation}
which, again not surprisingly, coincides with the advocated LD approach when $\xv_2 + \xv_3$ is a lattice point.

We conclude that the LZF and the LD approaches correspond to ML decoding under arbitrary interference constrained into some linear subspace, and
approximated MAP decoding taking into account the true discrete nature of the interference.

\section{Performance\label{sec:sim}}
It is true that the full rank criteria is almost-surely satisfied for channel coefficients drawn randomly from a continuous distribution.  However, in addition to rank criterion, the overall performance of IA with LZF highly depends on the orthogonality of the channel coefficient matrices $\Gb_1,\Gb_2,\Gb_3$. Notice that the channel matrices are ill-conditioned if two elements of the diagonal matrix $\Tb$ are close in value (and the rank criteria breaks if two elements of $\Tb$ are exactly the same). Further, the channel matrices  are also ill-conditioned if the absolute value of any of the channel coefficients deviates  from one, especially for larger $n$. The reason becomes clear by inspecting the exponential structure of the precoding matrices in \eqref{eq:pcoder1} and the diagonal channel ratio matrix $\Tb$ in \eqref{eq:pcoder4}; if any of the diagonal elements of $\Tb$ have an absolute value lower (or higher) than 1, the corresponding row in $\Vb_1$ exponentially tends to zero (or infinity) for large $n$. Thus, we expect a poor performance from linear alignment with LZF for dynamic channel amplitudes. 

\begin{figure}[t]
\includegraphics[width=1.05\columnwidth]{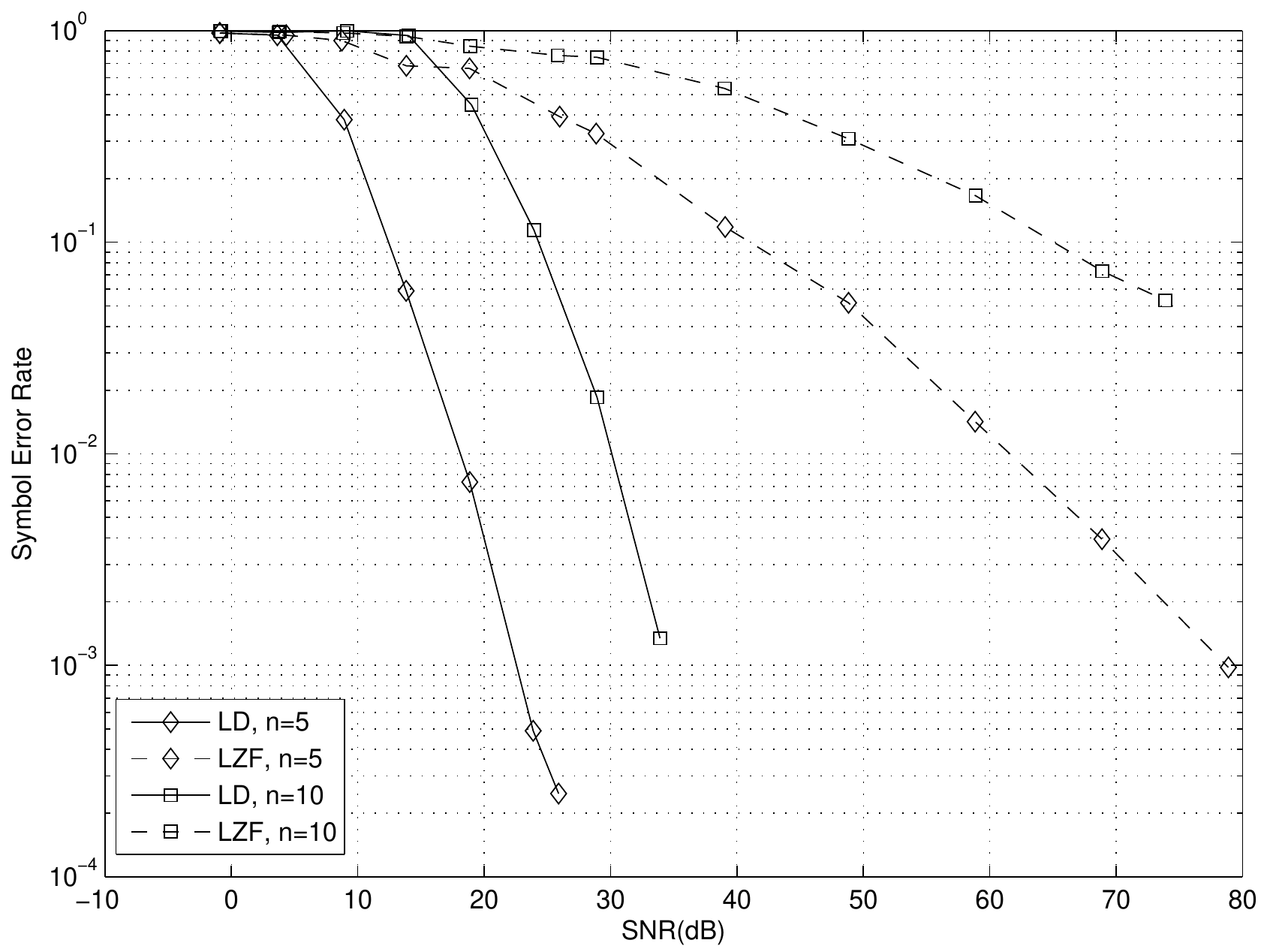}
\caption{Comparison between LZF and sphere lattice decoding (LD) for fixed channel amplitudes. The source constellation is 4-QAM. \label{fig:fig1}}
\end{figure}

\begin{figure}[t]
\includegraphics[width=1.05\columnwidth]{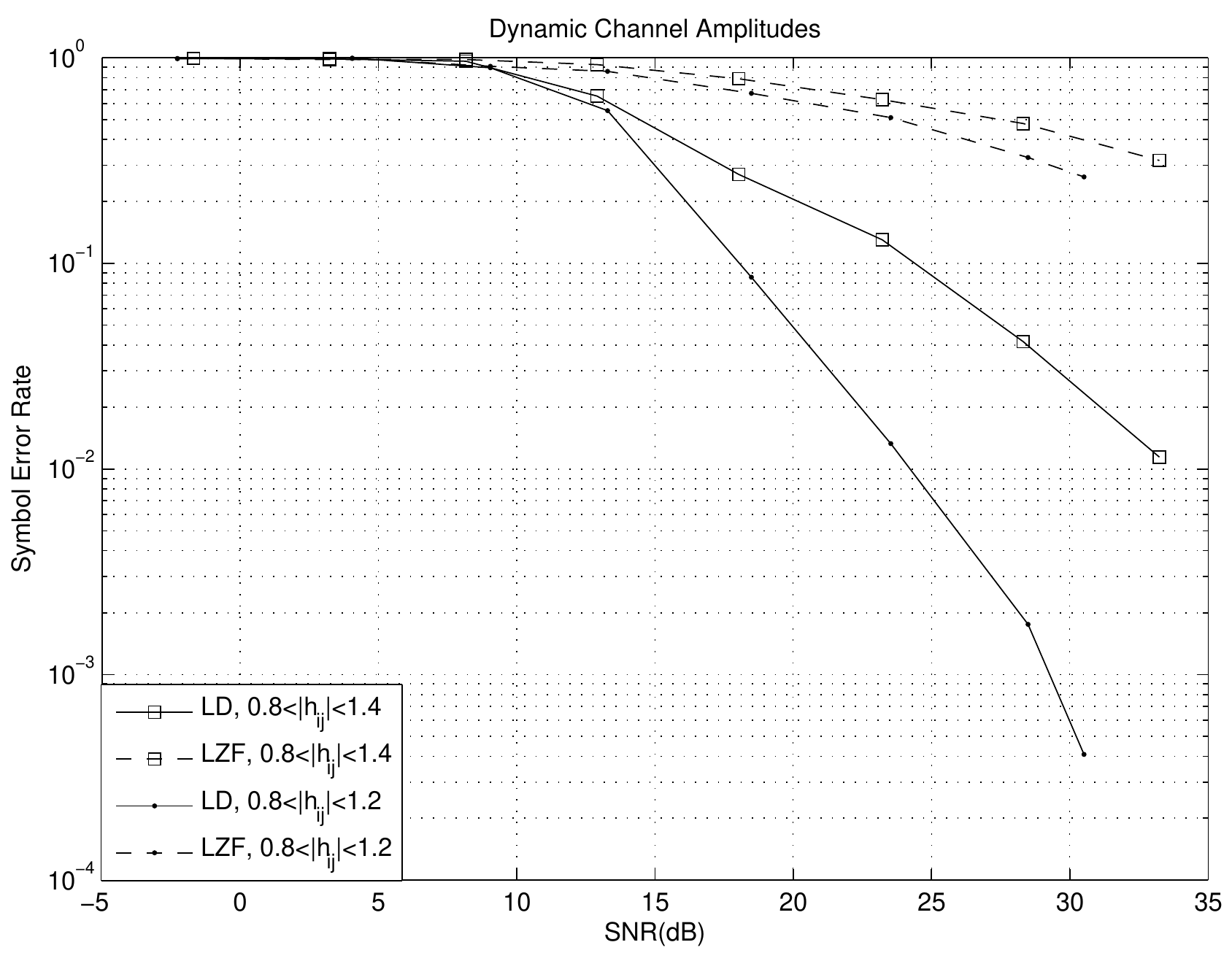}
\caption{Performance severely degrades with dynamic truncated channel amplitudes. Here, channel amplitudes  are truncated within $[0.8,1.2]$ and $[0.8,1.4]$. The source constellation is 4-QAM.  Blocklength is $N=11$, i.e., $n=5$. \label{fig:fig2}}
\end{figure}

Consider first a random channel where channel coefficients all have magnitude 1 with a random uniform phase, i.e., $H^t_{kl}=\exp\{j\phi_{kl}^t\}$ for i.i.d. $\phi_{kl}^t\sim\U(0,2\pi)$, $k,l=1,2,3,j=\sqrt{-1}$.
For this setup, Fig.~\ref{fig:fig1} shows a simulation comparison between filtering out the interference subspace (LZF), and a nonlinear strategy using Schnorr-Euchner sphere decoding strategy (LD) of \cite{damen_gamal_caire1}. In this figure, the vertical access represents symbol error rate, and the horizontal axis represents average SNR.  Here, SNR is defined as the ratio of average transmit power to noise, and is computed by averaging the realized transmit power (including the channel inversion) over a set of consecutive blocks for fixed constellation energy and received noise variance. We compare the linear alignment strategy with LZF, and the discrete alignment scheme with LD, for blocklengths $N=2n+1=21$ and $N=2n+1=11$. In all simulations, the source symbols are drawn from  a 4-QAM (QPSK) constellation.

Fig.~\ref{fig:fig1} shows that LD obtains a significant improvement over LZF. As shown in this figure, even with fixed channel amplitudes, performance of LZF is not quite acceptable. For SNR values as high as 80 dB, the symbol error rate is about $10^{-3}$ for $n=5$. However, discrete alignment with sphere decoding achieves a reasonable symbol error rate in the order of $10^{-3}$ for SNRs around 20--25dB.

Theoretically, larger blocklengths achieve higher degrees of freedom.  However, for the purpose of decoding at finite SNRs, larger blocklengths result in a higher symbol error rate, because of the skewed structure  of the equivalent MIMO channel. This can be seen by comparing the corresponding curves for both LZF and LD at $n=5$ and $n=10$ in Fig.~\ref{fig:fig1}. Recall that the structure of the equivalent MIMO channel matrix is exponential in $n$.

Fig.~\ref{fig:fig2} shows a similar comparison between linear alignment with LZF and discrete alignment with lattice decoding (LD), for truncated Gaussian coefficients with amplitudes within $[0.8,1.2]$ and $[0.8,1.4]$. In this simulation, channel coefficients are drawn from a complex circularly-symmetric normal distribution, and subchannels with a magnitude outside the desired interval are discarded.  In comparison to symbol-error rate performance with fixed channel amplitudes in Fig.~\ref{fig:fig1}, dynamic channel amplitudes significantly degrade the performance as shown in Fig.~\ref{fig:fig2}. In fact for larger block lengths, the advantage of LD over LZF quickly fades out with increasing channel amplitude dynamics.

It is intuitively predictable that smaller channel amplitudes directly result in worse performance.  However, Fig.~\ref{fig:fig2} shows that larger channel amplitudes are also not suitable for interference alignment and result in a significant performance loss. As shown in Fig.~\ref{fig:fig2}, LZF performs very poorly even at SNRs as high as 35dB for dynamic channel amplitudes within $[0.8,1.4]$.  However, with discrete alignment combined with LD, we obtain a reasonable symbol error rate of $10^{-2}$ (which combined with a powerful outer code could yield an acceptable overall bit-error rate) at SNR 35 dB. A noticeable improvement is further observed when channel amplitudes are truncated within $[0.8,1.2]$, which again, demonstrates the effect of channel dynamics on the overall performance. We conclude that the precoding strategy in \eqref{eq:pcoder1} and  \eqref{eq:pcoder4} is highly susceptible to power control and channel amplitude dynamics, and a tight power control strategy and carrier pairing is required to achieve a reasonable performance.

\section{Conclusion\label{sec:con}}
While the asymptotic degrees-of-freedom analysis is a powerful tool in understanding sophisticated communication systems, it provides little intuition about performance at finite SNRs.  In fact in many cases of interest, a one-dimensional (scalar) analysis suffices to compute asymptotic degrees of freedom, including the $K$-user interference channel \cite{motahari_gharan}, relay networks \cite{amaudruz_faragouli}, the two-user interference channel \cite{bresler_tse}, and many-to-one and one-to-many interference channels \cite{bresler_parekh_tse}.

In this paper, we examined the finite-SNR performance of linear subspace interference alignment strategy of \cite{cadambe_jafar} for frequency-selective interference channels with single transmit and receive antennas and i.i.d.\ continuous channel coefficients. A discrete alignment strategy is introduced that uses algebraic properties of lattices to transform the single-antenna interference channel to an equivalent MIMO system, for which efficient decoding strategies are well understood. While linear subspace alignment with zero forcing shows a poor performance at finite SNRs, discrete alignment along with sphere decoding strategy significantly boosts the performance.

We observed a particular performance sensitivity to channel amplitude dynamics.  
This suggests that to benefit from interference alignment, we need to employ a tight power control mechanism, combined with user selection, and carrier pairing in time and frequency dimensions to obtain near-constant channel amplitudes. 

 
The ergodic interference alignment strategy of \cite{nazer_IA} avoids exponential beamforming vectors and channel inverse operations, thus, could be more suitable for larger channel amplitude dynamics, e.g., Rayleigh fading. In ergodic interference alignment, the transmitting nodes, in a way, employ an extreme form of subcarrier pairing over the time horizon, essentially ``waiting for the right channel'' to align.  As a result of this opportunistic approach, however, ergodic alignment suffers from an average decoding delay that grows exponentially as $O(\exp(B K^2))$, where $B$ is the number of quantization bits used to represent the channel coefficients (which also grows with the operating SNR).
On the other hand, the CJ alignment strategy ignores the inherent randomness of channel coefficients and employs a deterministic precoding strategy that works for a fixed deterministic blocklength, though the penalty in power could be large. 
The next step to further improve the performance could be using opportunistic strategies to control channel variations to obtain new interference alignment strategies in between the CJ scheme and ergodic alignment.

\bibliographystyle{IEEEtran}
\bibliography{IEEEabrv,reference}

\end{document}